\pdfoutput=1

\documentclass[12pt]{article}
\usepackage{graphicx}
\usepackage{multirow}

\def\napoli{Universit\`a del Piemonte Orientale and INFN, Alessandria, Italy}

\def\Title#1{\begin{center} {\Large #1 } \end{center}}
\def\Author#1{\begin{center}{ \sc #1} \end{center}}
\def\Address#1{\begin{center}{ \it #1} \end{center}}

\newenvironment{Abstract}{\begin{quotation}  }{\end{quotation}}
\newenvironment{Presented}{\begin{quotation} \begin{center} 
             PRESENTED AT\end{center}\bigskip 
      \begin{center}\begin{large}}{\end{large}\end{center} \end{quotation}}

\def\beq{\begin{equation}}
\def\eeq#1{\label{#1}\end{equation}}
\def\eeqn{\end{equation}}

\def\beqa{\begin{eqnarray}}
\def\eeqa#1{\label{#1}\end{eqnarray}}
\def\eeqan{\end{eqnarray}}

\let\bar=\overbar

\def\Dslash{\not{\hbox{\kern-4pt $D$}}}
\def\dslash{\not{\hbox{\kern-2pt $\del$}}}

\def\msb{{\bar{\ssstyle M \kern -1pt S}}}

\begin{document}
\begin{titlepage}

\vfill
\Title{Muon bundles from cosmic rays with ALICE}
\vfill
\Author{Mario Sitta}
\begin{center}
on behalf of the ALICE Collaboration
\end{center}
\Address{\napoli}
\vfill
\begin{Abstract}
ALICE, a general purpose experiment designed to investigate nucleus-nucleus
collisions at the CERN Large Hadron Collider (LHC), has also been used to
detect atmospheric muons produced by cosmic-ray interactions in the atmosphere.

In this contribution the analysis of the multiplicity distribution of the
atmospheric muons detected by ALICE between 2010 and 2013 is presented, along
with a comparison with Monte Carlo simulations. Special emphasis is given to
the study of high-multiplicity events, i.e.\ those containing more than 100
reconstructed muons. Such high-multiplicity events demand primary cosmic rays
with energy above $10^{16}$~eV. The frequency of these events can be
successfully described by assuming a heavy mass composition of primary cosmic
rays in this energy range, using the most recent interaction models to describe
the development of the air shower resulting from the primary interaction.
\end{Abstract}
\vfill
\begin{Presented}
Presented at EDS Blois 2017, Prague, \\ Czech Republic, June 26-30, 2017
\end{Presented}
\vfill
\end{titlepage}
\def\thefootnote{\fnsymbol{footnote}}
\setcounter{footnote}{0}

\section{Introduction}
The use of collider detectors to study the atmospheric muons was pioneered by
the LEP experiments ALEPH, DELPHI and L3. All results reported by the LEP
experiments were consistent with the standard hadronic interaction models,
except the observation of high-multiplicity muon-bundle events: even under the
assumption of the highest measured flux and a pure-iron spectrum, the Monte
Carlo models of that time failed to describe the rate of those
high-multiplicity events \cite{Ref:ALEPHRes},\cite{Ref:DELPHIRes}.

ALICE (A Large Ion Collider Experiment) \cite{Ref:AliceJINST} is a
general-purpose, heavy-ion detector at the CERN LHC. It focuses on the study of
the properties of the Quark-Gluon Plasma (QGP) created in strongly interacting
matter at extreme energy densities in high energy nucleus-nucleus collisions.
Besides the study of heavy-ion physics, since 2009 ALICE developed a cosmic-ray
physics program, which exploits the excellent tracking capabilities of the
Time Projection Chamber (TPC) to detect and reconstruct the muons produced by
the cosmic radiation interacting with the atmosphere.

ALICE is located at Point 2 of the LHC tunnel, 52~m underground and with 28~m
of overburden rock. This depth of rock completely absorbs the electromagnetic
and hadronic components of the cosmic-ray induced air shower, and poses an
energy threshold of 16~GeV for vertical muons \cite{Ref:EnerThr}. The ALICE
solenoid magnet (maximum strength 0.5~T) houses the central-barrel detectors,
where different techniques are exploited to detect and reconstruct all
particles coming out from the primary interaction vertex. A forward muon arm,
is located on one side outside the solenoid magnet. A detailed description of
the ALICE detector is given in \cite{Ref:AliceJINST}.

For the cosmic-ray data taking three detectors were used for the trigger.
The Alice COsmic Ray DEtector (ACORDE) is an array of 60 scintillator modules
located on the three top octants of the ALICE magnet, covering 10\% of their
surface. The Silicon Pixel Detector (SPD) forms the two innermost coaxial
cylinders of the Inner Tracking System (ITS) around the beam pipe and it is
composed by 10M pixels segmented into 120 modules; for cosmic studies it can
provide a trigger. The Time Of Flight (TOF) detector is a cylindrical array of
1638 Multi-gap Resistive Plate Chamber pads arranged in 18 sectors, surrounding
the TPC; the trigger configuration used for the present analysis requires a
coincidence between a signal in a sector of the upper part and a signal in a
sector in the opposite lower part.

The ALICE Time Projection Chamber (TPC) is the largest detector of this type,
and is the main ALICE tracking device. It has an inner radius of 80~cm, an
outer radius of 280~cm and a total length of 500~cm along the beam axis. Filled
with a mixture of Ne-CO$_2$-N$_2$\/, it is read out by multi-wire proportional
chambers at both end caps. The total area to detect cosmic muons is about
26~m$^2$\/, however after applying a minimum length requirement to the
reconstructed muon tracks the effective area reduces to about 17~m$^2$\/.

\section{Data selection and reconstruction}
The data used for the cosmic-ray analysis presented here \cite{Ref:Cosmic} were
collected between 2010 and 2013 during periods without circulating beams in the
LHC. Cosmic-ray data were recorded with a logical OR of at least two out of the
three aforementioned triggers, depending on the run period. The integrated live
time amounts to 30.8 days, during which $\sim$22.6M events with at least 1
reconstructed muon in the TPC were accumulated. A multi-muon event is defined
as an event with at least 5 muons; the data sample contains 7487 multi-muon
events.

The TPC tracking algorithm was designed to reconstruct tracks coming out from
the interaction region, working inwards from the outer radius. As a consequence
a cosmic muon crossing the TPC gets reconstructed as two separate tracks,
called {\it up} and {\it down}\/, belonging to the two halves of the TPC
cylinder. A specific algorithm was worked out to match the two track segments
as a single one. Real and Monte Carlo events of different multiplicities were
used to optimize the parameters of this matching algorithm. Moreover, to avoid
possible reconstruction inaccuracies associated with the most inclined tracks,
the zenith angle of all accepted tracks was restricted to $0 < \theta <
50^\circ$\/.

TPC tracks were required to consist of at least 50 clusters (out of a maximum
of 159) and, in events where the magnetic field was on, to have a minimum
momentum of 0.5~GeV/$c$\/, to eliminate all possible background from $e^\pm$\/.
For multi-muon events a parallelism condition was also applied, which requires
the angular difference in space between two tracks to satisfy $\cos\Delta\psi >
0.990$\/. Finally, to match up and down tracks a maximum distance of closest
approach of 3~cm in the TPC middle plane was imposed. A muon reconstructed with
two TPC tracks (up and down) is called a {\it matched muon}\/; a track
satisfying all selection criteria but the maximum distance is still accepted as
muon candidate but flagged as {\it single-track muon}\/. Most single-track
muons are particles crossing the TPC near the edges where part of their
trajectory may fall outside the sensitive volume.
\vspace{-0.2cm}
\section{Muon multiplicity distribution}
The main topic related to the cosmic-ray physics investigated by ALICE is the
study of the muon-multiplicity distribution (MMD). The MMD, obtained from the
whole data sample and corrected for trigger efficiency, shows a smooth
distribution up to a muon multiplicity of $\sim$~70 and then 5 events with a
multiplicity greater than 100. Events with $N_{\mu} > 100$ are denoted High
Muon Multiplicity (HMM) events.

In order to understand the MMD, simulated events equivalent to 30.8 days of
live time were generated using CORSIKA \cite{Ref:CORSIKA} as event generator
and QGSJET \cite{Ref:QGSJET} for the hadronic interaction model. Two samples,
pure $p$ (representing a light composition) and pure Fe (representing an
extremely heavy composition), were generated. The primary cosmic-ray energy
was restricted to the interval $10^{14} < E < 10^{18}$~eV following the usual
power law energy spectrum $E^{-\gamma}$\/, with a spectral index $\gamma = 2.7$
below the knee ($E_k = 3\times10^{15}$~eV) and $\gamma = 3.0$ above. The total
all-particle absolute flux was extracted from \cite{Ref:AllFlux}. For each
shower the core was randomly scattered at surface level over an area of
205$\times$205~m$^2$ centered above ALICE.

\begin{figure}[htb]
     \centering
        \vspace{-0.4cm}
        {\includegraphics[scale=0.570]{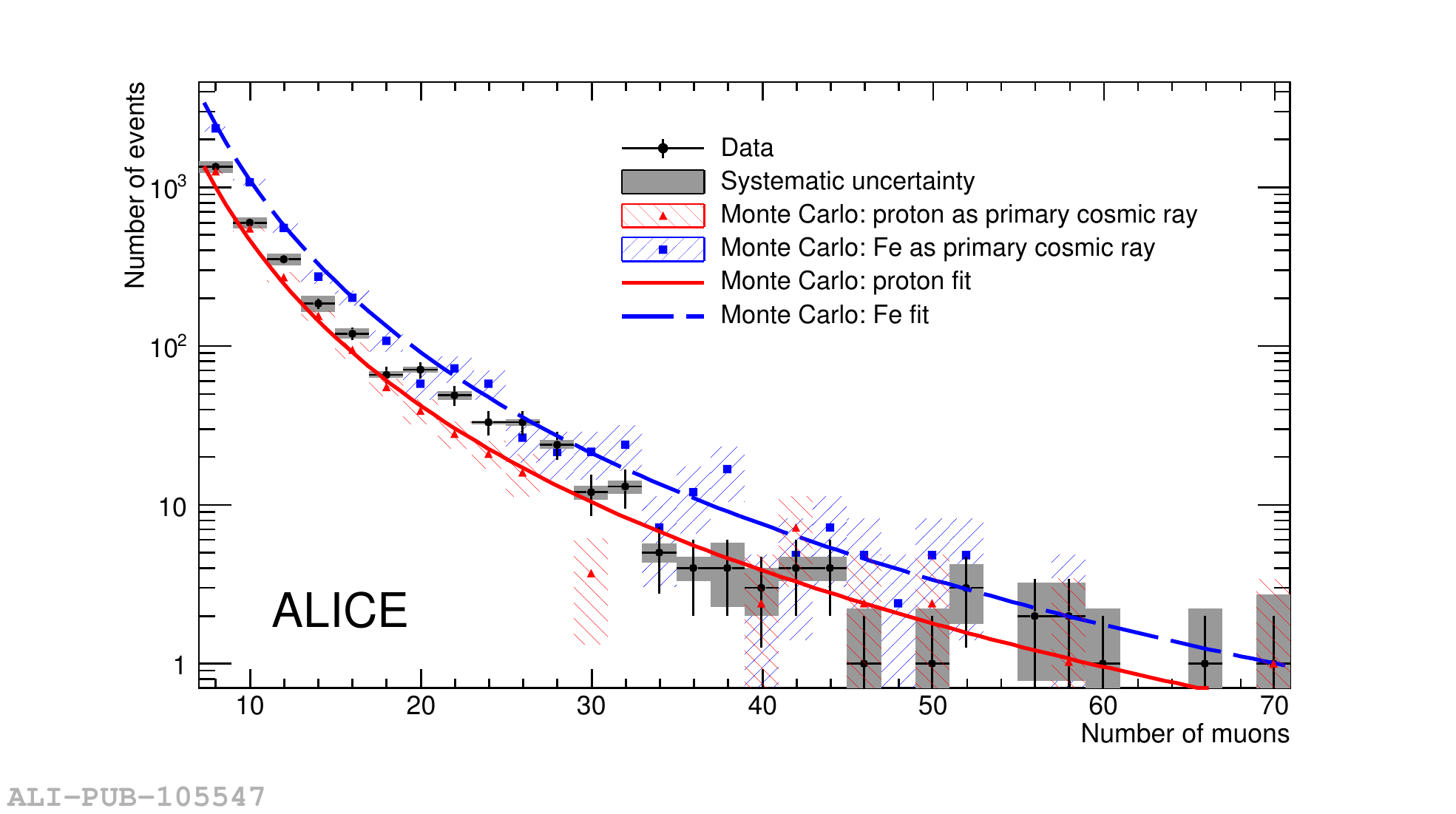}}
        \vspace{-0.4cm}
        \caption{\it Measured MMD compared with values and fits obtained from
simulations with $p$ and Fe primaries. Figure taken from \cite{Ref:Cosmic}.}
        \label{Fig:ALICEDataMC}
\end{figure}

The comparison between the MMD in the range $7 < N_{\mu} < 70$ and the
simulated distribution fitted with a power-law function is shown in Fig.\
\ref{Fig:ALICEDataMC} \cite{Ref:Cosmic}, where errors are shown separately
(statistical and systematic) for data, and summed in quadrature for Monte
Carlo. Below $N_{\mu} \sim 30$ the data, as expected, are between the pure $p$
composition (approaching it at low multiplicity) and the pure Fe composition
(at higher multiplicity). Above $\sim 30$ the low statistics does not allow us
to draw any firm conclusion, though the experimental points, considering their
errors, are inside the region limited by the $p$ and Fe curves.

\vspace{-0.7cm}
\section{High muon multiplicity events}
In 30.8 live days 5 HMM events were recorded, corresponding to a rate of
$1.9\times10^{-6}$~Hz. The highest multiplicity cosmic-muon event reconstructed
in the TPC was found to contain 276 muons. To estimate the rate of these
events, while limiting the fluctuations in the number of HMM simulated events,
a live time equivalent of 1 year was simulated. The full simulation was
restricted to primaries with $10^{16} < E < 10^{18}$~eV, since only primaries
within this energy range contribute to these events. To further reduce the
statistical fluctuations, four additional simulations were performed, reusing
the same EAS sample and randomly varying the shower core in the
205$\times$205~m$^2$ area. Given that the TPC acceptance is some 3000 times
smaller, this ensures that the samples are statistically independent. By
averaging the 5 samples the number of HMM events in 1 year is estimated while
reducing the statistical fluctuations.

In Tab.\ \ref{Tab:HMMEvents} \cite{Ref:Cosmic} the results from Monte Carlo
simulations are compared with data. The rate of HMM events can be well
reproduced by the latest interaction models and a primary flux extrapolated
from the direct measurements at 1~TeV. A pure Fe primary composition seems in
closer agreement with the measured rate, though the large uncertainty of the
latter prevents a definite conclusion about the origin of these events. This is
consistent with the fact that HMM events stem from primaries with energy $>
10^{16}$~eV, where the composition is expected to be dominated by heavier
elements.

\begin{table}[h]
\centering
\vspace{-0.3cm}
\caption{\it Comparison of the HMM event rate between data and Monte Carlo
\cite{Ref:Cosmic}.}
\vskip 0.1 in
\begin{tabular}{|l|c|c|c|c|c|} \hline
\multirow{3}{*}{HMM events}&
\multicolumn{2}{|c|}{CORSIKA 6990}&\multicolumn{2}{|c|}{CORSIKA 7350}&\\
&\multicolumn{2}{|c|}{QGSJET II-03}&\multicolumn{2}{|c|}{QGSJET II-04}&Data\\
\cline{2-5}
& $p$ & Fe & $p$ & Fe & \\
\hline
\hline
Period [days per event] & 15.5 & 8.6 & 11.6 & 6.0 & 6.2 \\
Rate [$\times 10^{-6}$ Hz] & 0.8 & 1.3 & 1.0 & 1.9 & 1.9 \\
Uncertainty (sys+stat) (\%) & 25 & 25 & 22 & 28 & 49 \\
\hline
\end{tabular}
\label{Tab:HMMEvents}
\end{table}
\vspace{-0.4cm}
\section{Conclusions}
In 2010--2013 the ALICE Experiment collected 30.8 live days of cosmic-ray data.
The MMD distribution at low and intermediate multiplicities is well reproduced
by Monte Carlo simulations using CORSIKA with the QGSJET model. The
measurements by ALICE presented here suggest a mixed ion primary cosmic-ray
composition with an average mass increasing with energy. In the same period 5
HMM events were recorded. The observed rate is consistent with the predictions
of the most recent CORSIKA and QGSJET models using a pure Fe primary
composition and energies $> 10^{16}$~eV. For the first time the rate of HMM
events has been well reproduced using conventional hadronic interaction models
and a primary flux extrapolated from low energies.

\end{document}